 \documentclass[final,3p,times,twocolumn]{elsarticle}

\usepackage{graphicx}
\usepackage{wrapfig}

\usepackage{epsfig,bm}
\usepackage{graphics}
\usepackage{amsmath}
\usepackage{mathrsfs}
\usepackage[normalem]{ulem}  
\usepackage[dvips]{color} 

\renewcommand\sout{\bgroup \color{red} \ULdepth=-.5ex \ULset}

\biboptions{sort&compress}

\begin{document}

\begin{frontmatter}


\title{An in-Medium Heavy-Quark Potential from the $Q\bar{Q}$ Free Energy}

\author{Shuai~Y.~F.~Liu and Ralf Rapp}

\address{Cyclotron Institute and Department of Physics and Astronomy, 
Texas A$\&$M University, College Station, Texas 77843-3366, USA
}
\begin{abstract}
We investigate the problem of extracting a static potential between a quark 
and its antiquark in a quark-gluon plasma (QGP) from lattice-QCD computations 
of the singlet free energy, $F_{Q\bar{Q}}(r)$. We utilize the thermodynamic 
$T$-matrix formalism to calculate the free energy from an underlying 
potential ansatz resummed in ladder approximation. Imaginary parts of both 
$Q\bar Q$ potential-type and single-quark selfenergies are included as 
estimated from earlier results of the $T$-matrix approach. We find that the 
imaginary parts, and in particular their (low-) energy dependence, induce 
marked deviations of the (real part of the) potential from the calculated 
free energy. When fitting lattice results of the latter, the extracted 
potential is characterized by significant long-range contributions from 
remnants of the confining force. We briefly discuss consequences of this 
feature for the heavy-quark transport coefficient in the QGP. 
\end{abstract}


\end{frontmatter}



\section{Introduction}
Heavy-flavor (HF) particles, containing charm or bottom quarks, play a
central role in the analysis of strongly interacting matter as produced 
in ultralrelativistic heavy-ion collisions 
(URHICs)~\cite{Rapp:2009my}. 
The nuclear modification factor and elliptic flow of $D$ and $B$ mesons 
(as well as their semileptonic decay leptons) are key measures to 
establish and quantify the strong-coupling nature of the QCD 
medium~\cite{Adare:2006nq,ALICE:2012ab,Adamczyk:2014uip}, while 
the production patterns of charmonia and 
bottomonia~\cite{Adare:2006ns,Abelev:2012rv,Chatrchyan:2012lxa,Adamczyk:2013poh}
are believed to encode information on its deconfinement properties. 
For both observables, a good understanding of the in-medium 
interactions of the heavy quarks and quarkonia is mandatory. 

Lattice-QCD (lQCD) computations allow to study the properties of heavy-quark 
(HQ) systems in the quark-gluon plasma (QGP) from first principle. In 
particular, euclidean correlation functions of quarkonia, HQ susceptibilities 
and static $Q\bar Q$ free energies have been computed with good 
precision~\cite{Bazavov:2009us,Skullerud:2014sla}. A non-perturbative 
$T$-matrix approach (``Brueckner theory") for quarkonia and heavy quarks in 
the QGP has been developed to interpret the lQCD 
results~\cite{Cabrera:2006wh,Riek:2010fk}, and the calculated spectral 
functions and transport coefficients have been applied to both 
hidden~\cite{Zhao:2010nk} and open HF observables~\cite{vanHees:2007me} 
in URHICs. The main theoretical uncertainty in this and other approaches 
based on the Schr\"odinger 
equation~\cite{shuryak2004toward,Mocsy:2005qw,wong2005heavy,Beraudo:2007ky,Lee:2013dca} 
remains the definition of the driving kernel or potential. Thus far, it 
has been bracketed by using either the temperature dependent free or 
internal energies (or combinations thereof), 
which, roughly speaking, led to a rather weakly or strongly coupled systems, 
respectively. 

In the present paper we address this uncertainty by utilizing the thermodynamic 
$T$-matrix to derive an expression for the singlet HQ free energy, $F_{Q \bar{Q}}(r)$. 
Previous work to extract the HQ potential has been done both nonperturbatively
from lQCD using an effective Schr\"{o}dinger equation and potential for the
Wilson loop~\cite{Rothkopf:2011db,Burnier:2012az,Bazavov:2014kva,Burnier:2014ssa},
and within weak-coupling schemes~\cite{Laine:2006ns,Beraudo:2007ky,Brambilla:2008cx}.
Here, we take a diagramatic approach by defining the potential within a
field-theoretic scattering equation, which leads to a slightly different
Schr\"odinger (or Dyson) equation than in previous work.
Starting from the finite-temperature Bethe-Salpeter equation (BSE) 
for a static quark and antiquark in a mixed coordinate-frequency 
representation, we resum the ladders into a closed form expression which 
naturally serves as the definition of the in-medium driving kernel. The 
pertinent spectral function can then be straightforwardly 
related to the HQ free energy. A key element in evaluating the resulting 
expression is the energy dependence of the spectral function, in particular 
of imaginary parts induced by both one- and two-body correlations, i.e., 
single-quark self-energies and $Q\bar Q$ potential-type contributions. The 
former have been calculated before within the selfconsistent $T$-matrix 
formalism~\cite{Riek:2010py}, while for the latter we take guidance from 
perturbative calculations~\cite{Laine:2006ns,Beraudo:2007ky,Brambilla:2008cx}.
The real part of the potential will be taken from a field-theoretic ansatz for 
a screened Cornell potential~\cite{Megias:2007pq}, where four parameters 
characterize its strength and screening properties. This set-up will then 
be deployed to conduct a fit to lQCD data for the static color-singlet free energy.

\section{Static $Q\bar{Q}$ Free Energy in Ladder Approximation}
\label{se_bulk}
Our objective in this section is to relate the static correlation function 
of (infinitely massive) quark and antiquark to the HQ free energy, ${F_{Q\bar{Q}}(r)}$
\cite{McLerran:1981pb,kapusta2006finite,Beraudo:2007ky,Beraudo:2010tw}, and 
evaluate it in ladder approximation (BSE or $T$-matrix scheme). Toward this 
end we first formally carry out the ladder-diagram resummation of the BSE in 
the imaginary-time (Matsubara) formalism, and then reduce the BSE to the 
$T$-matrix (potential approximation).

The one-particle propagator in the infinite-mass 
limit is given by~\cite{Beraudo:2007ky}
\begin{align}
&G^{(1)}  \left(v_n ,\textbf{r}'\right)
=\int \frac{d^3\textbf{p}'}{(2\pi)^3}e^{i\textbf{p}'\cdot\textbf{r}'}
\frac{1}{i v_n-\varepsilon _{\textbf{p}'}-\Sigma^{(1)} \left(v_n,\textbf{p}'\right)}
\nonumber\\
&\approx\int \frac{d^3\textbf{p}'}{(2\pi)^3}e^{i\textbf{p}'\cdot\textbf{r}'}
\frac{1}{i v_n-M-\Sigma^{(1)} \left(v_n\right)}\equiv\delta\left(\textbf{r}'\right)
\tilde{G}^{(1)}\left(v_n\right) \ ,
\label{eq_Green1F}	
\end{align}
where the $\delta$-function implies that the particle never moves; $\Sigma^{(1)}$
denotes the single-particle selfenergy and $M$ is its (heavy) mass;  
$v_n=(2n+1)\pi /\beta$ is a discrete fermionic Matsubara frequency ($n$ integer, 
$T=1/\beta$: temperature), adopting the conventions of Ref.~\cite{fetter2003quantum} 
with $\hbar=1$. The two-particle Green function inherits the $\delta$-function 
structure~\cite{Beraudo:2007ky}\footnote{Color and spin indices are 
suppressed.}:
\begin{align}
&G^>(-i\tau ,\textbf{r}_1,\textbf{r}_2|\textbf{r}_1',\textbf{r}_2')\nonumber\\
	&\equiv\delta\left(\textbf{r}_1-\textbf{r}_1'\right)
\delta\left(\textbf{r}_2-\textbf{r}_2'\right)\tilde{G}^>(-i\tau, r)\label{eq_G} \ ,
\end{align}
where $r=|\textbf{r}_1-\textbf{r}_2|$. Here and in the following the ``tilde" 
notation is used to denote the reduced Green function without the spatial 
$\delta$-functions. The static $Q\bar{Q}$ free energy, $F_{Q\bar{Q}}$, can be 
defined in terms of the $Q\bar Q$ Green function as~\cite{Beraudo:2007ky}
\begin{align}
 &F_{Q\bar{Q}}(r)=-\frac{1}{\beta }\ln \left(\tilde{G}^>\left(-i\beta ,r\right)
\right) \ .
\label{eq_defineF}
\end{align}
The task in the following is to calculate $\tilde{G}^>(-i\tau, r)$ in 
Eq.~(\ref{eq_G}) within a BSE (or potential model) in the Matsubara formalism.
In a ``mixed" frequency-coordinate representation it is given by
\begin{eqnarray}
\tilde{T}\left(z_\lambda,v_n,v_m|r\right)=
K\left(v_n-v_m,r\right) -\frac{1}{\beta}\sum_{k} K\left(v_n-v_k,r\right)
\nonumber\\
\times G_0^{(2)}\left(z_\lambda-v_k,v_k|r\right)\tilde{T}(z_\lambda,v_k,v_{m}|r) 
\quad
\label{eq_Tnm}	
\end{eqnarray}
where $ K\left(v_n-v_m,r\right) $ is the interaction kernel and 
\begin{equation}
G_0^{(2)}(z_\lambda-v_n,v_n|r) \equiv \tilde{G}^{(1)})(z_\lambda-v_n)
\tilde{G}^{(1)}(v_n)
\end{equation}
the uncorrelated 
two-particle propagator. The external frequency $z_\lambda$ is conserved 
because of imaginary-time translation invariance. The full two-particle 
Green function takes the form~\cite{kraeft1986quantum,kadanoff1962quantum}
\begin{align}
&\tilde{G}\left(z_\lambda,r\right)=
\tilde{G}_0^{(2)}(z_\lambda)
\nonumber\\
&+\frac{1}{\beta^2}\sum_{k,l}\tilde{G}_0^{(2)}(z_\lambda-v_k,v_k)
\tilde{T}(z_\lambda,v_k,v_l|r)\tilde{G}_0^{(2)}(z_\lambda-v_l,v_l) \ ,
\label{eq_Green4PsF}
\end{align}
where we defined $\tilde{G}_0^{(2)}(z_\lambda)\equiv
-\frac{1}{\beta}\sum_{k}\tilde{G}_0^{(2)}(z_\lambda-v_k,v_k)$, and 
likewise for $\tilde{G}(z_\lambda,r)$. The discrete energy dependencies 
and summations render Eq.~(\ref{eq_Tnm}) of matrix type, 
$T_{nm} = K_{nm} + \sum_k [KG_0^{(2)}]_{nk} T_{km}$,  
which is amenable to a formal solution via matrix inversion,   
$T_{nm} = \sum_k [{1}-KG_0^{(2)}]_{nk}^{-1} K_{km}$, which, however,
is not very practical. 

To proceed, we adopt the a potential approximation, i.e., neglecting
energy transfers in the interaction kernel so that 
$K(v_n-v_k,r)\to V(z_\lambda,r)$; we keep a possible dependence on the 
external energy parameter $z_\lambda$, in analogy to what has been 
done, e.g., in the description of electromagnetic 
plasmas~\cite{zimmermann1978dynamical} where a ``dynamical screening" 
can induce such a dependence.
With this approximation, the Matsubara summations in Eq.~(\ref{eq_Tnm}) 
decouple and one can resum the geometric series to find
\begin{align}
&\tilde{T}\left(z_\lambda,v_n,v_m|r\right)=
\nonumber\\
&V(z_\lambda,r) - \frac{1}{\beta}\sum_k V(z_\lambda,r) 
\tilde{G}_0^{(2)}(z_\lambda-v_k,v_k)
\tilde{T}(z_\lambda,v_k,v_m|r)
\nonumber\\
& \quad \quad =\frac{V(z_\lambda,r)}{1-V(z_\lambda,r)
	\tilde{G}_0^{(2)}(z_\lambda)} 
\equiv\tilde{T}(z_\lambda|r)
\label{eq_Tz}
\end{align}
The corresponding full two-particle Green function can also be written
in compact form as  
\begin{equation}
\tilde{G}(z_\lambda,r)=\frac{1}{\left[\tilde{G}_0^{(2)}(z_\lambda)\right]^{-1}
-V(z_\lambda,r)} \ . 
\label{eq_G2}	
\end{equation}
The above expresson is a key formula of our derivation, as it can serve as the 
definition of the interaction kernel in the $T$-matrix formalism
It slightly differs from what has been adopted in previous 
approaches~\cite{Beraudo:2007ky,Brambilla:2008cx,Rothkopf:2011db,Burnier:2014ssa}, as can be
seen from inspecting the resulting Dyson equation for the meson spectral function. 

Next we use the analyticity of the Green function to obtain the correlation 
function in imaginary-time via a spectral representation,
\begin{align}
&\tilde{G}^>\left(-i\tau ,r\right)
=\int\limits_{-\infty }^{\infty
}dE'\sigma_V \left(E',r\right)\frac{e^{E'(\beta-\tau)}}{e^{\beta E'}-1} \ ,
\label{eq_GreenTF}
\end{align}
with the spectral function corresponding to Eq.~(\ref{eq_G2}),  
\begin{equation}
\sigma_V (E,r)= -\frac{1}{\pi }\tilde{G}_I\left(E+i\epsilon |r\right) \ . 
\label{eq_SpecF}
\end{equation}
We use the subscript ``$V$" to indicate that the spectral function is 
calculated in potential approximation. Here and in the following, the 
subscripts "$R$" and ``$I$" are short-hand notations for the real and 
imaginary part of a given quantity. 
Since we work in the static limit ($M\to \infty$), we redefine the spectral
function relative to the $Q\bar Q$ threshold by a shift $E=E'-2M$,
i.e., $\bar{\sigma}_V(E,r) \equiv \sigma_V(E+2M,r)$,
which further simplifies Eq.~(\ref{eq_GreenTF}) to    
\begin{align}
&\bar{G}^>(-i\tau ,r)\equiv\frac{\tilde{G}^>(-i\tau,r)}{e^{-\tau (2M)}}=
\int\limits_{-\infty }^{\infty} dE\bar{\sigma}_V(E,r) {\rm e}^{-\tau E}
\label{eq_GreenTR}
\end{align} 
Combining Eqs. (\ref{eq_defineF}), (\ref{eq_SpecF}) and (\ref{eq_GreenTR}), 
we arrive at the final expression for $F_{Q \bar{Q}}(r)$ within the $T$-matrix
approach,  
 \begin{align}
 &F_{Q\bar{Q}}=-\frac{1}{\beta }\ln \left(\int\limits_{-\infty }^{\infty}
 dE \bar{\sigma}_V(E,r) {\rm e}^{-\beta E}\right)
 \label{eq_defineFCon}
 \end{align}
In the following section, in the spirit of a variational approach, 
we will fit a physically motivated ansatz for the potential to lQCD
data of the in-medium free energy.

\section{Potential Extraction from the Free Energy}
\label{sec_extract}
The two-particle Green function in medium receives contributions from the 
(medium-modified) interaction kernel, $V$, and from the single-particle 
selfenergies, $\Sigma^{(1)}$, arising from individual interactions of each 
particle with the heat bath which figure into $G_2^{(0)}$. Both are, in 
principle, complex quantities and are contained in our $T$-matrix expression 
for the full two-particle Green function, Eq.~(\ref{eq_G2}); 
we write the pertinent spectral function as  
\begin{align}
&\bar{\sigma}_V(E,r)= -\frac{1}{\pi }\tilde{G}_I(E+2M+i\epsilon |r) \approx
\nonumber\\
&\frac{1}{\pi}\frac{-\left[V_I(E,r)+\Sigma^{(2)}_I(E)\right]}
{\left[E-V_R(E,r)-\Sigma^{(2)}_R(E)\right]^2+
\left[V_I(E,r)+\Sigma^{(2)}_I(E)\right]^2} \ , 
\label{eq_SpecFC}
\end{align}
where we use $\Sigma^{(2)}$ to denote the two-particle selfenergy that arises from 
the uncorrelated single-particle selfenergies~\cite{kraeft1986quantum,Riek:2010py}.
We now make concrete ans\"atze for the potential and selfenergies in terms of 
a field-theoretic model for the in-medium Cornell potential~\cite{Megias:2007pq}
and previous results from the $T$-matrix formalism~\cite{Riek:2010py}. The former
leads to the following real parts of potential and uncorrelated selfenergy, 
\begin{align}
&V_R(E,r)=-\frac{4}{3}\alpha _s\frac{e^{-m_Dr}}{r}-\sigma \frac{e^{-m_sr}}{m_s}
\label{eq_VR}
\\
&\Sigma^{(2)}_R(E)=-\frac{4}{3}\alpha _sm_D+\sigma\frac{1}{m_s}
\label{eq_SelfER}
\ ,
\end{align}
corresponding to the $r$-dependent and $r$-independent terms of the screened Cornell 
potential, respectively (additional contributions from the uncorrelated 2-particle 
propagator are small~\cite{Riek:2010py} and therefore neglected in the present 
exploratory calculation). Their strengths are characterized by the strong coupling, 
$\alpha_s=g^2/(4\pi)$, and string tension $\sigma$, along with corresponding
screening masses $m_D$ and $m_s$, respectively. 
As in our previous work~\cite{Riek:2010fk}, we utilize the latter as 
independent fit parameters, where, for simplicity, we choose 
a linear ans\"atze, $m_{D,s}=c_{D,s}gT$ with two constant coefficients,
$c_{D,s}$ (which turn out to be close of order one). 
The imaginary part of the uncorrelated 2-particle selfenergy, $\Sigma^{(2)}_I$,
follows from the folding integral of the single-particle propagators which 
contain the respective in-medium selfenergies. In principle, this quantity 
follows automatically from a selfconsistent calculation, once the potential is 
specified, as was carried out in Ref.~\cite{Riek:2010py}. For our exploratory
study in the present work, we do not enforce selfconsistency but instead  
employ the energy dependence of $\Sigma^{(1)}_I$ as found in Ref.~\cite{Riek:2010py} 
(see Fig.~3 in there) to compute $\Sigma^{(2)}_I$. It turns out that the 
latter can be well represented by Gaussian form, 
\begin{equation}
\Sigma^{(2)}_I(E )=D\exp \left(-\frac{(E-\frac{\Sigma^{(2)}_R}{2})^2}{2(CT)^2}\right)
\frac{4}{3}\alpha _sT \ . 
\label{eq_SigI}
\end{equation}
With $C\simeq1.6$ we can reproduce the energy dependence of the single-particle 
width of Ref.~\cite{Riek:2010py}, being very similar for both internal and 
free energies as underlying potential. The strength parameter $D$, however,
which we define relative to the perturbative value, $\frac{4}{3}\alpha_s T$, 
markedly depends on the underlying potential, varying by a factor of 2-3 
between free and internal energies; as mentioned above, we here use it as a 
fit parameter.

Finally, we have to specify the imaginary part of the potential; its 
magnitude and $r$ dependence has been determined previously in the weak
coupling limit as $V_I^{\rm pert}=\frac{4}{3}\alpha _sT\Phi \left(m_Dr\right)$
\cite{Laine:2006ns,Beraudo:2007ky,Brambilla:2008cx}, where $\Phi(x)$ is
a smoothly varying function which vanishes for $x$=0 and approaches 1 for
large $x$. However, in our definition of the potential, we have identified 
the non-zero part at infinite distance with the uncorrelated selfenergy. 
Upon subtracting this part and again allowing for a nonperturbative 
enhancement with the same stength parameter $D$ and energy dependence 
as in Eq.~(\ref{eq_SigI}) we arrive at the form  
\begin{equation}
V_I(E,r)=D\exp \left(-\frac{(E-\frac{\Sigma^{(2)}_R}{2})^2}{2(CT)^2}\right)
\frac{4}{3}\alpha _sT [\Phi(B m_Dr)-1] \ . 
\label{eq_VI}
\end{equation}
To accommodate the possibility that the imaginary part of the string 
interaction can have a different $r$ dependence than given by $\Phi$ (albeit 
with the same asymptotic limits), we rescaled its argument by a constant $B$ 
(which turns out to be close to one).  
We note that nontrivial energy dependencies in both imaginary parts, 
$V_I$ and $\Sigma_I^{(2)}$, are crucial to ensure convergence of the
integration in Eq.(\ref{eq_defineFCon}). In this sense, the potential is 
required to go beyond the static approximation. 


As an example of our method we perform a fit to a set of free energies computed in 
lattice QCD with $N_f=2+1$ dynamical flavors~\cite{kaczmarek2007screening}, for 
temperatures $T$=1.2, 1.7, 2.2 $T_c$, with $T_c$=196\,MeV~\cite{kaczmarek2007screening}.
The following parameter values were used: The Coulomb Debye mass is $m_D = 1.22 g T$, 
close to the expected HTL value, while the string interaction is screened less with 
$m_s = 0.8 g T$; for the string tension itself a constant value of  
$\sigma = 1.58$\,GeV/fm turns out to be sufficient, which is larger than in vacuum but 
significantly smaller than in previous fits to the 
in-medium free energy~\cite{Megias:2007pq,Riek:2010fk}. For the strong coupling 
we utilize a weak temperature running, $\alpha_s=0.288-0.05(\frac{T}{1.2 Tc}-1)$, 
decreasing from 0.288 at 1.2$T_c$ to 0.238 at 2.4\,$T_c$. For the dimensionless 
coefficients we have $B=0.73$, which delays the approach of $\Phi$ to its 
asymptotic value of one, and $C$=2.07 in Eq.~(\ref{eq_SigI}), which produces slightly
longer tails in the energy dependence of the HQ selfenergies than in the microscopic
calculations of Ref.~\cite{Riek:2010py}.     

\begin{figure*}[!hbt]
	\centering
	\includegraphics[width=0.33\textwidth]{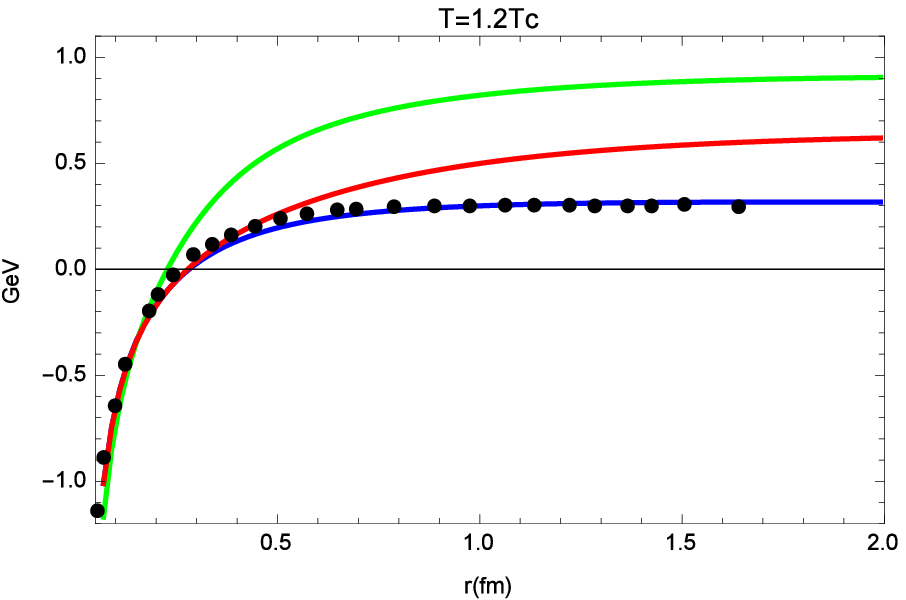}
	\includegraphics[width=0.33\textwidth]{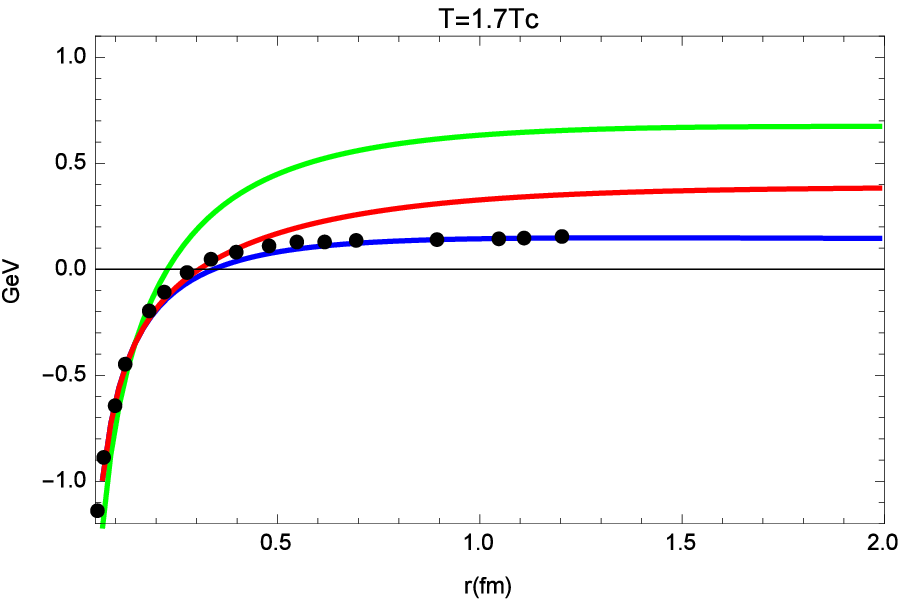}
	\includegraphics[width=0.33\textwidth]{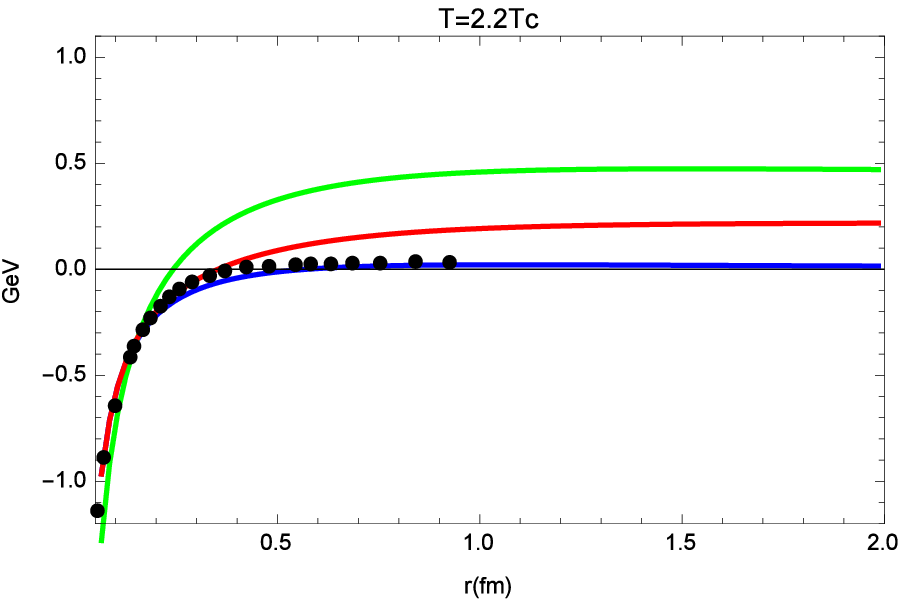}
\caption{Our fit (blue lines) to the color-singlet $Q\bar Q$ free energies computed
in lattice QCD (black dots)~\cite{kaczmarek2007screening} at 3 different temperatures,
$T$=1.2, 1.7, 2.2 $T_c$ in the left, middle and right panel, respectively. 
The red lines are the real parts of the underlying potential, while the green lines 
represent the correpsonding internal energies.
}
\label{fig_freeEfit}
\end{figure*}

The resulting fit to the lQCD free energies is displayed in Fig.~\ref{fig_freeEfit}, 
showing good agreement. Reproducing the temperature dependence essentially ensures 
that the internal energies, $U=F+TS$ with $S=-dF/dT$, also agree with the lattice 
data.  The underlying two-body potential turns out to lie significantly above 
$F_{Q\bar{Q}}$, but below $U_{Q\bar{Q}}$, at intermediate and long distances. 
However, the potential differs from both free and internal energy in that its 
derivative, characterizing the inter-quark force, is larger than for both 
$F_{Q\bar{Q}}$ and $U_{Q\bar{Q}}$, in a region around $r\simeq 1$\,fm. 
This is a remnant of the confining string term in the hot medium, caused by 
the relatively small screening mass, $m_s$. Even though this force is not so large, 
it has a rather long range, encompassing a relatively large volume. In the heavy-light
quark sector, this enables heavy quarks to effectively couple to more particles
in the heat bath, relative to a short-range force. Our preliminary calculations of 
the HQ diffusion coefficient indeed indicate that the latter is comparable to 
earlier calculations using the internal energy~\cite{Riek:2010fk}, which has a stronger 
force but over a shorter range. In this sense, the potential obtained in the present
study may still imply a strongly coupled QGP.

The above exmaple of a fit is not unique, mostly due to uncertainties in the 
imaginary parts, which in the future will have to be calculated in a selfconsistent 
scheme as laid out in Ref.~\cite{Riek:2010py}.\footnote{The imaginary parts implicit
in our fit are comparable to what was found before with the internal energy as 
potential, which indicates that they are in the expected ballpark of the selfconsistent 
solution.} However, several qualitative, model-independent observations can be made.
First, if the imaginary parts, $(V+\Sigma^{(2)})_I(E)$, are taken to zero, the spectral 
function $\sigma_V$ becomes a $\delta$ function in energy and the potential in the 
denominator
is forced to become the free energy, thus recovering the weak-coupling analysis of
Ref.~\cite{Beraudo:2007ky}. This underlines the key role of the imaginary parts of 
both potential and single-quark selfenergy in raising the potential above the free
energy, i.e., rendering a stronger force. The $E$-depedence of $(V+\Sigma^{(2)})_I(E)$ 
(the ``width of the width"), encoded in the width parameter $C$, also plays an important
role, which reiterates the need for microscopic calculations, such as the $T$-matrix
approach. The dependence on the  real part is  more robust and ``model independent" 
in the sense that it can be largely determined by the small distance behavior of free 
energy where $\Phi(x)$ forces the imaginary part to become small. 
Let us also comment on the magnitude of the string tension exceeding its vacuum value
by about 60\%. This appears to be a rather generic feature within the screened Cornell 
potential ansatz. In previous works utilizing this 
ansatz~\cite{Megias:2007pq,Riek:2010fk}, where the strength of the string term was 
additionally allowed to develop a temperature dependence, fits to the free energy
preferred string tensions of up to a factor 3-4 times the vacuum, albeit with a much
stronger screening than in our present potential fit. As a result, the in-medium
free energy never exceeds the vacuum potential (as in our present work). Similar
features are also found in applications of the Dyson-Schwinger approach to quarkonia 
at finite $T$~\cite{Qin:2014dqa}. A possible microscopic mechanism for an enhanced 
string tension could be the release of magnetic monopole in the near-$T_c$
region, as suggested, e.g., in Ref.~\cite{Liao:2008vj}.

\section{Conclusions and Outlook}
Utilizing a thermodynamic $T$-matrix approach, we have derived an expression for the 
static heavy-quark free energy that directly relates to the underlying interaction
kernel, resummed in ladder approximation. The such defined kernel serves as our 
definition of an in-medium potential, $V_{Q \bar{Q}}(r)$, in the context of a diagrammatic 
many-body approach. Key elements of this framework are a non-trivial energy dependence 
of the pertinent $Q\bar Q$ spectral function, induced by the imaginary parts of both 
one- and two-body type, i.e., in the single-quark selfenergies and the potential. We
have applied this set-up to fit lattice-QCD ``data" for the color-singlet free energy 
in the spirit of a variational scheme for the underlying potential. To facilitate this,  
we made an ansatz for the real part of the potential in terms of a screened Cornell 
potential, augmented with an imaginary part motivated by perturbative studies but 
with variable magnitude. We additionally accounted for complex heavy-quark self-energies 
with an energy dependence motivated by earlier selfconsistent implementations of the 
many-body theory. The resummations of the interaction kernel, with a nonperturbative
string term, and of the selfenergies are required in a regime of strong coupling, as 
expected for the QGP at temperatures not too far above $T_c$.
Indeed, our resulting potential significantly deviates from the weak coupling limit 
(in which case it would be close to the free energy), and exhibits a rather long-range 
remnant of the confining force surviving well above $T_c$. The associated imaginary parts 
are large, not inconsistent with what was obtained earlier using the internal energy as
a potential approximation. First estimates indicate that the pertinent HQ diffusion 
coefficient is around $D_s(2\pi T) \simeq 3-5$, suggestive for a strong coupling regime. 

Several further investigations are in order to scrutinize our initial estimates within
the $T$-matrix approach. First, the calculation needs to be carried out selfconsistently,
by computing the heavy-quark selfenergies from the underlying $T$-matrices and reinserting
the former in the latter~\cite{Riek:2010py}. Second, the free energies should be updated
with the most recent lattice-QCD results. Third, quarkonium correlators, HQ suscpetibilities 
and transport coefficients should be computed and systematically compared to lattice data. 
The role of possible three-body correlations may need to be addressed. If the theoretical 
framework passes these tests, applications to heavy-quark and quarkonium phenomenology are 
warranted.


\section*{Acknowledgements}
This work was supported by the U.S. National Science Foundation under Grant No. PHY-1306359.


\bibliographystyle{elsarticle-num}
\bibliography{refc}

\begin{thebibliography}{10}
\expandafter\ifx\csname url\endcsname\relax
  \def\url#1{\texttt{#1}}\fi
\expandafter\ifx\csname urlprefix\endcsname\relax\def\urlprefix{URL }\fi
\expandafter\ifx\csname href\endcsname\relax
  \def\href#1#2{#2} \def\path#1{#1}\fi

\bibitem{Rapp:2009my}
R.~Rapp, H.~van Hees, in: R.~Hwa, X.~N. Wang (Eds.), Quark-Gluon Plasma 4,
  World Scientific, Singapore, 2010, p. 111.
\newblock \href {http://arxiv.org/abs/0903.1096 [hep-ph]}
  {\path{arXiv:0903.1096 [hep-ph]}}.

\bibitem{Adare:2006nq}
A.~Adare, et~al., PHENIX Collaboration Phys. Rev. Lett. 98 (2007) 172301.

\bibitem{ALICE:2012ab}
B.~Abelev, et~al., PHENIX Collaboration JHEP 1209 (2012) 112.

\bibitem{Adamczyk:2014uip}
L.~Adamczyk, et~al., STAR Collaboration Phys. Rev. Lett. 113 (2014) 142301.

\bibitem{Adare:2006ns}
A.~Adare, et~al., PHENIX Collaboratio Phys. Rev. Lett. 98 (2007) 232301.

\bibitem{Abelev:2012rv}
B.~Abelev, et~al., ALICE Collaboration Phys. Rev. Lett. 109 (2012) 072301.

\bibitem{Chatrchyan:2012lxa}
S.~Chatrchyan, et~al., CMS Collaboration Phys. Rev. Lett. 109 (2012) 222301.

\bibitem{Adamczyk:2013poh}
L.~Adamczyk, et~al., STAR Collaboration Phys. Lett. B 735 (2014) 127.

\bibitem{Bazavov:2009us}
A.~Bazavov, P.~Petreczky, A.~Velytsky, in: R.~Hwa, X.~N. Wang (Eds.),
  Quark-Gluon Plasma 4, World Scientific, Singapore, 2010, p.~61.
\newblock \href {http://arxiv.org/abs/0904.1748 [hep-ph]}
  {\path{arXiv:0904.1748 [hep-ph]}}.

\bibitem{Skullerud:2014sla}
J.-I. Skullerud, G.~Aarts, C.~Allton, A.~Amato, Y.~Burnier, et~al., {\,}\href
  {http://arxiv.org/abs/1501.00018 [nucl-th]} {\path{arXiv:1501.00018
  [nucl-th]}}.

\bibitem{Cabrera:2006wh}
D.~Cabrera, R.~Rapp, Phys. Rev. D 76 (2007) 114506.

\bibitem{Riek:2010fk}
F.~Riek, R.~Rapp, Phys. Rev. C 82 (2010) 035201.

\bibitem{Zhao:2010nk}
X.~Zhao, R.~Rapp, Phys. Rev. C 82 (2010) 064905.

\bibitem{vanHees:2007me}
H.~van Hees, M.~Mannarelli, V.~Greco, R.~Rapp, Phys. Rev. Lett. 100 (2008)
  192301.

\bibitem{shuryak2004toward}
E.~V. Shuryak, I.~Zahed, Phys. Rev. D 70 (2004) 054507.

\bibitem{Mocsy:2005qw}
A.~Mocsy, P.~Petreczky, Phys. Rev. D 73 (2006) 074007.

\bibitem{wong2005heavy}
C.-Y. Wong, Phys. Rev. C 72 (2005) 034906.

\bibitem{Beraudo:2007ky}
A.~Beraudo, J.-P. Blaizot, C.~Ratti, Nucl. Phys. A 806 (2008) 312.

\bibitem{Lee:2013dca}
S.~H. Lee, K.~Morita, T.~Song, C.~M. Ko, Phys. Rev. D 89 (2014) 094015.

\bibitem{Rothkopf:2011db}
A.~Rothkopf, T.~Hatsuda, S.~Sasaki, Phys. Rev. Lett. 108 (2012) 162001.

\bibitem{Burnier:2012az}
Y.~Burnier, A.~Rothkopf, Phys. Rev. D 86 (2012) 051503.

\bibitem{Bazavov:2014kva}
A.~Bazavov, Y.~Burnier, P.~Petreczky, Nucl. Phys. A 932 (2014) 117.

\bibitem{Burnier:2014ssa}
Y.~Burnier, O.~Kaczmarek, A.~Rothkopf, {\,}\href
  {http://arxiv.org/abs/1410.2546 [hep-lat]} {\path{arXiv:1410.2546
  [hep-lat]}}.

\bibitem{Laine:2006ns}
M.~Laine, O.~Philipsen, P.~Romatschke, M.~Tassler, JHEP 0703 (2007) 054.

\bibitem{Brambilla:2008cx}
N.~Brambilla, J.~Ghiglieri, A.~Vairo, P.~Petreczky, Phys. Rev. D 78 (2008)
  014017.

\bibitem{Riek:2010py}
F.~Riek, R.~Rapp, New J. Phys. 13 (2011) 045007.

\bibitem{Megias:2007pq}
E.~Megias, E.~Ruiz~Arriola, L.~Salcedo, Phys. Rev. D 75 (2007) 105019.

\bibitem{McLerran:1981pb}
L.~D. McLerran, B.~Svetitsky, Phys. Rev. D 24 (1981) 450.

\bibitem{kapusta2006finite}
J.~I. Kapusta, C.~Gale, Finite-temperature field theory: Principles and
  applications, Cambridge Univ. Press, 2006.

\bibitem{Beraudo:2010tw}
A.~Beraudo, J.~Blaizot, P.~Faccioli, G.~Garberoglio, Nucl. Phys. A 846 (2010)
  104.

\bibitem{fetter2003quantum}
A.~L. Fetter, J.~D. Walecka, Quantum theory of many-particle systems, Courier
  Dover Publications, 2003.

\bibitem{kraeft1986quantum}
W.-D. Kraeft, D.~Kremp, W.~Ebeling, G.~R{\"o}pke, Quantum statistics of charged
  particle systems, Springer, 1986.

\bibitem{kadanoff1962quantum}
L.~P. Kadanoff, G.~A. Baym, Quantum statistical mechanics, Benjamin, 1962.

\bibitem{zimmermann1978dynamical}
R.~Zimmermann, K.~Kilimann, W.~Kraeft, D.~Kremp, G.~R{\"o}pke, physica status
  solidi (b) 90 (1978) 175.

\bibitem{kaczmarek2007screening}
O.~Kaczmarek, PoS CPOD07 (2007) 043.

\bibitem{Qin:2014dqa}
S.-x. Qin, D.~H. Rischke, Phys. Lett. B 734 (2014) 157.

\bibitem{Liao:2008vj}
J.~Liao, E.~Shuryak, Phys. Rev. D 82 (2010) 094007.

\end{thebibliography}

\end{document}